\documentclass[10pt,conference]{IEEEtran}
\IEEEoverridecommandlockouts
\usepackage{cite}
\usepackage{amsmath,amssymb,amsfonts}
\usepackage{algorithmic}
\usepackage{algorithm}
\usepackage{graphicx}
\usepackage[caption=false,font=footnotesize,labelfont=rm,textfont=rm]{subfig}
\usepackage{textcomp}
\usepackage{xcolor}
\def\BibTeX{{\rm B\kern-.05em{\sc i\kern-.025em b}\kern-.08em
    T\kern-.1667em\lower.7ex\hbox{E}\kern-.125emX}}
\usepackage[left=1.62cm,right=1.62cm,top=1.9cm]{geometry}
\begin{document}
\title{Secure Full-Duplex Communication via Movable Antennas \\
\thanks{This work was supported in part by Science and Technology Project of Guangzhou under Grants 202206010118 and 2023B04J0011 and in part by National Natural Science Foundation of China under Grant 62171006. The calculations were supported by High-Performance Computing Platform of Peking University.  \emph{(Corresponding author: Zijian Zhou.)}}
}

\author{\IEEEauthorblockN{Jingze Ding, Zijian Zhou, Chenbo Wang, Wenyao Li, Lifeng Lin, and Bingli Jiao}
\IEEEauthorblockA{School of Electronics, Peking University, Beijing, China\\
djz@stu.pku.edu.cn; \{zjzhou1008, wcb15\}@pku.edu.cn; liwenyao@stu.pku.edu.cn; \{linlifeng, jiaobl\}@pku.edu.cn}
}

\maketitle

\begin{abstract}
This paper investigates physical layer security (PLS) in a movable antenna (MA)-assisted full-duplex (FD) system. In this system, an FD base station (BS) with multiple MAs for transmission and reception provides services for an uplink (UL) user and a downlink (DL) user. Each user operates in half-duplex (HD) mode and is equipped with a single fixed-position antenna (FPA), in the presence of a single-FPA eavesdropper (Eve).  To ensure secure communication, artificial noise (AN) is transmitted to obstruct the interception of Eve.  The objective of this paper is to maximize the sum secrecy rate (SSR) of the UL and DL users by jointly optimizing the beamformers of the BS and the positions of MAs.  This paper also proposes an alternating optimization (AO) method to address the non-convex problem, which decomposes the optimization problem into three subproblems and solves them iteratively.  Simulation results demonstrate a significant performance gain in the SSR achieved by the proposed scheme compared to the benchmark schemes.
\end{abstract}


\section{Introduction}
The broadcast nature of wireless communications exposes data transmission to interception and unauthorized access.  Therefore, implementing physical layer security (PLS) \cite{ref1} becomes crucial to safeguard sensitive information and ensure transmission privacy.  The multiple-antenna technique \cite{ref2} is an effective strategy for enhancing security, as it can capitalize on the spatial degrees of freedom (DoFs) to improve the legitimate channel while degrading the eavesdropping channel.  However, traditional multiple-antenna systems typically utilize fixed-position antennas (FPAs), which restricts their ability to further exploit channel variations, especially in cases with a limited number of antennas.

To overcome this limitation, movable antennas (MAs) \cite{ref_MA_Modeling, ref_MA_MIMO, ref_MA_Multiuser, ref_MA_Beamforming1,multi_ding} offer a practical solution.  The MA enables flexible movement through a driver, such as a step motor along a slide track, which is also known as fluid antenna system with other implementation ways in antenna positioning \cite{Historical}.  Some studies, using MAs, have shown improvements in secure systems over FPAs \cite{ref_MA_Secure1, ref_MA_Secure2, ref_MA_Secure3}.  For instance, \cite{ref_MA_Secure1} investigated the use of linear MA array to achieve secure wireless communications, maximizing the achievable secrecy rate by jointly designing the transmit beamforming and antennas' positions.  \cite{ref_MA_Secure2} and \cite{ref_MA_Secure3} focused on secure systems aided by multiple MAs.  The former utilized alternating optimization and gradient descent to tackle the non-convex optimization problem, while the latter developed a block coordinate descent and majorization-minimization-based algorithm to solve a similar issue.

Nevertheless, the terminals in these works operate in half-duplex (HD) mode, which cannot secure both uplink (UL) and downlink (DL) transmissions simultaneously. On the other hand, the growing data demands faster communication rates using limited spectrum resources, posing a challenge for HD systems.  Therefore, full-duplex (FD) communication \cite{ref10, ref11, ref12}, which enables simultaneous transmission and reception by overlapping signals in the same time-frequency block, has been employed.  FD systems rely on self-interference (SI) cancellation (SIC) techniques to achieve this capability and protect both UL and DL transmissions, in which artificial noise (AN) is usually superimposed onto DL information.

With these considerations, this paper aims to explore secure communication through an FD base station (BS) equipped with separate transmit and receive MAs.  In this system, the BS can dynamically alter the positions of MAs and adjust the beamformers for transmission and reception.  The single-FPA legitimate users are served by the BS, in the presence of a single-FPA eavesdropper (Eve).  Under this, the main contributions of this paper are summarized as follows.  First, we, for the first time, propose the system model and formulate an optimization problem to maximize the sum secrecy rate (SSR) of UL and DL transmissions using multiple MAs.  Second, we develop an alternating optimization (AO) method to iteratively solve three subproblems derived from the original optimization problem.  Last, simulation results are presented to demonstrate the superior performance of our proposed system.

\textit{Notations}: $a$, $\mathbf{a}$, and $\mathbf{A}$ represent a scalar, a vector, and a matrix, respectively. $\left(\cdot\right) ^\mathrm{T}$, $\left(\cdot\right) ^\mathrm{H}$, $\left\| \cdot \right\|_2$, and $\mathrm{Tr} \left( \cdot \right) $ denote transpose, conjugate transpose, Euclidean norm, and trace, respectively.  ${\left[  x  \right]^ + }$ indicates that $x$ is a non-negative value, i.e., $\max \left\{ { x ,0} \right\}$. $\odot$ is the Hadamard product.  $\mathbb{C}^{M \times N}$ and $\mathbb{R}^{M \times N}$ are the sets of $M \times N$-dimensional complex and real matrices, respectively.  $\mathbf{I}_K$ stands for the identity matrix of order $K$.  $\mathcal{CN}\left( 0, \mathbf{\Lambda}\right) $ denotes the circularly symmetric complex Gaussian (CSCG) distribution with mean zero and covariance matrix $\mathbf{\Lambda}$.

\section{System Model}
\begin{figure}[!t]
	\centering
	\includegraphics[width=0.7\columnwidth]{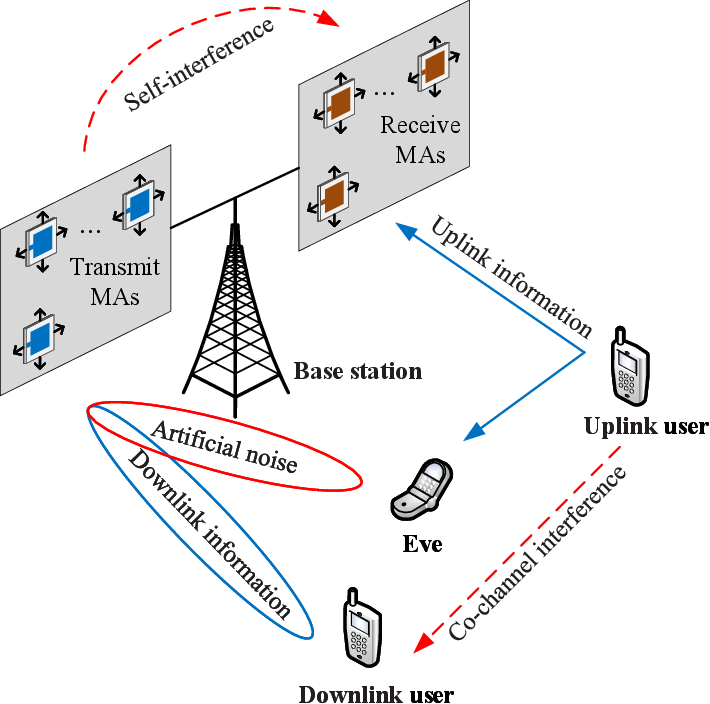}
	\caption{An illustration for the proposed communication system.}
	\label{fig1}
\end{figure}

As shown in Fig.\;\ref{fig1}, we consider a communication system with an MA-assisted FD BS to serve an HD UL user and an HD DL user, in the presence of an Eve\footnote{The scenario with multiple users and Eves has already been considered in our future work \cite{MA_Sec_Ding1}.}.  The BS is equipped with $N_\mathrm{t}$ transmit MAs and $N_\mathrm{r}$ receive MAs\footnote{The practical implementation methods of MA can be found in \cite{implementation}.}, and the users and the Eve each have a single FPA.  The positions of the ${n_\mathrm{t}}$-th transmit MA and the ${n_\mathrm{r}}$-th receive MA can be described by the vectors ${\mathbf{t}_{{n_\mathrm{t}}}} = {\left[ {{x_{\mathrm{t}}^{n_\mathrm{t}}},{y_{\mathrm{t}}^{n_\mathrm{t}}}} \right]^\mathrm{T}} \in {\mathcal{C}_\mathrm{t}}$ for $1 \le {n_\mathrm{t}} \le {N_\mathrm{t}}$ and ${\mathbf{r}_{{n_\mathrm{r}}}} = {\left[ {{x_{\mathrm{r}}^{n_\mathrm{r}}},{y_{\mathrm{r}}^{n_\mathrm{r}}}} \right]^\mathrm{T}} \in {\mathcal{C}_\mathrm{r}}$ for $1 \le {n_\mathrm{r}} \le {N_\mathrm{r}}$, where ${\mathcal{C}_\mathrm{t}}$ and ${\mathcal{C}_\mathrm{r}}$ represent the moving regions.

Let us define the collections of $N_\mathrm{t}$ transmit MAs and $N_\mathrm{r}$ receive MAs as $\tilde {\mathbf{t}} = {\left[ {\mathbf{t}_1^\mathrm{T},\mathbf{t}_2^\mathrm{T}, \cdots ,\mathbf{t}_{{N_\mathrm{t}}}^\mathrm{T}} \right]^\mathrm{T}} \in {\mathbb{R}^{2{N_\mathrm{t}} \times 1}}$ and $\tilde {\mathbf{r}} = {\left[ {\mathbf{r}_1^\mathrm{T},\mathbf{r}_2^\mathrm{T}, \cdots ,\mathbf{r}_{{N_\mathrm{r}}}^\mathrm{T}} \right]^\mathrm{T}} \in {\mathbb{R}^{2{N_\mathrm{r}} \times 1}}$, respectively.  For the MA-assisted system, the channel response can be written as the function of MAs' positions \cite{ref_MA_Modeling}. Thus, the SI channel of the BS, the channel from the BS to the DL user, the channel from the BS to the Eve, and the channel from the UL user to the BS are denoted as ${\mathbf{H}_\mathrm{SI}} \left( {\tilde{ \mathbf{t}}}, {\tilde{ \mathbf{r}}} \right) \in {\mathbb{C}^{{N_\mathrm{r}} \times {N_\mathrm{t}}}}$, ${\mathbf{h}_\mathrm{BD}}\left( {\tilde{ \mathbf{t}}} \right)  \in {\mathbb{C}^{{N_\mathrm{t}} \times 1}}$, ${\mathbf{h}_\mathrm{BE}} \left( {\tilde{ \mathbf{t}}} \right) \in {\mathbb{C}^{{N_\mathrm{t}} \times 1}}$, and ${\mathbf{h}_\mathrm{UB}} \left( {\tilde{ \mathbf{r}}} \right) \in {\mathbb{C}^{{N_\mathrm{r}} \times 1}}$, respectively.  Besides, we denote the channel from the UL user to the DL user as ${h_\mathrm{UD}} \in {\mathbb{C}^{1 \times 1}}$ and to the Eve as ${h_\mathrm{UE}} \in {\mathbb{C}^{1 \times 1}}$.  The channel state information (CSI) of the above channels is assumed to be perfectly known at the BS\footnote{Although obtaining perfect CSI is challenging, this assumption allows us to establish an upper bound for performance analysis.}.

To ensure security, the BS is to generate the AN vector $\mathbf{v} \in {\mathbb{C}^{{N_\mathrm{t}} \times 1}}$ and transmit it with DL information to interfere with the interception of the Eve.  Therefore, the DL signal can be expressed as $\mathbf{x} = \mathbf{w}s_\mathrm{D} + \mathbf{v}$, where $s_\mathrm{D}$ is the information for the DL user with zero mean and normalized power, and $\mathbf{w} \in {\mathbb{C}^{{N_\mathrm{t}} \times 1}}$ is the corresponding beamforming vector.

Then, the received signals at the BS, the DL user, and the Eve can be respectively expressed as
\begin{align}
	{y_\mathrm{U}} &= \underbrace {\mathbf{w}_\mathrm{r}^\mathrm{H}{\mathbf{h}_\mathrm{UB}\left( {\tilde{ \mathbf{r}}} \right)}\sqrt {P_\mathrm{U}} {s_\mathrm{U}}}_\text{UL information} + \underbrace {\mathbf{w}_\mathrm{r}^\mathrm{H} \sqrt \rho {\mathbf{H}_\mathrm{SI}\left( {\tilde{ \mathbf{t}}}, {\tilde{ \mathbf{r}}} \right)}\mathbf{x}}_\text{SI} + \mathbf{w}_\mathrm{r}^\mathrm{H}\mathbf{n}_\mathrm{B} , \\
	{y_\mathrm{D}} &= \underbrace{{\mathbf{h}_\mathrm{BD}^\mathrm{H}\left( {\tilde{ \mathbf{t}}} \right)} \mathbf{x}}_\text{DL signal} + \underbrace{{h_\mathrm{UD}} \sqrt{P_\mathrm{U}} {s_\mathrm{U}}}_\text{Co-channel interference} + n_\mathrm{D} ,\\
	{y_\mathrm{E}} &= \underbrace{{\mathbf{h}_\mathrm{BE}^\mathrm{H}\left( {\tilde{ \mathbf{t}}} \right)} \mathbf{x}}_\text{DL signal} + \underbrace{{h_\mathrm{UE}} \sqrt{P_\mathrm{U}} {s_\mathrm{U}}}_\text{UL information} + n_\mathrm{E} ,
\end{align}
where ${s_\mathrm{U}}$ is the UL information with zero mean and normalized power and ${P_\mathrm{U}}$ is the transmitted power of the UL user.  $\rho$ represents the SIC coefficient.  The vector $\mathbf{w}_\mathrm{r}\in {\mathbb{C}^{{N_\mathrm{r}} \times 1}}$ is the receive beamformer. ${\mathbf{n}_\mathrm{B}} \sim \mathcal{CN}\left( {\mathbf{0},\sigma _\mathrm{B}^2{\mathbf{I}_{{N_\mathrm{r}}}}} \right)$, ${n_\mathrm{D}} \sim \mathcal{CN}\left( {0,\sigma _\mathrm{D}^2} \right)$, and ${n_\mathrm{E}} \sim \mathcal{CN}\left( {0,\sigma _\mathrm{E}^2} \right)$ stand for the additive white Gaussian noise at the BS, the DL user, and the Eve, respectively.

\subsection{Channel Model}
According to the field-response based channel model \cite{ref_MA_Modeling, ref_MA_MIMO,ref_MA_Secure2}, the channel responses for ${\mathbf{H}_\mathrm{SI}} \left( {\tilde{ \mathbf{t}}}, {\tilde{ \mathbf{r}}} \right)$, ${\mathbf{h}_\mathrm{UB}} \left( {\tilde{ \mathbf{r}}} \right)$, ${\mathbf{h}_\mathrm{BD}} \left( {\tilde{ \mathbf{t}}} \right)$, and ${\mathbf{h}_\mathrm{BE}} \left( {\tilde{ \mathbf{t}}} \right)$ are established as follows.

\subsubsection{MIMO Channel for ${\mathbf{H}_\mathrm{SI}} \left( {\tilde{ \mathbf{t}}}, {\tilde{ \mathbf{r}}} \right)$}
The channel between the transmit and receive MAs at the BS is a multiple input multiple output (MIMO) channel.  We denote the number of transmit and receive paths as ${{L}_\mathrm{t,SI}}$ and ${{L}_\mathrm{r,SI}}$, respectively. The normalized wave vector of the $l_\mathrm{t,SI}$-th ($1 \le l_\mathrm{t,SI} \le L_\mathrm{t,SI}$) transmit path can be represented by ${\mathbf{s}^{l_\mathrm{t,SI}}} = \left[ {\sin \theta _\mathrm{t,SI}^{l_\mathrm{t,SI}} \cos \phi _\mathrm{t,SI}^{l_\mathrm{t,SI}}, \cos \theta _\mathrm{t,SI}^{l_\mathrm{t,SI}}} \right]$, where $\theta _\mathrm{t,SI}^{l_\mathrm{t,SI}}, \phi _\mathrm{t,SI}^{l_\mathrm{t,SI}} \in \left[ {0,\pi} \right]$ are the elevation and azimuth angles of departure, respectively. Then, the phase difference of the $l_\mathrm{t,SI}$-th transmit path between the $n_\mathrm{t}$-th transmit MA's position ${\mathbf{t}_{{n_\mathrm{t}}}}$ and the origin of the transmit region ${\mathbf{o}_\mathrm{t}} = {\left[ {0,0} \right]^\mathrm{T}}$ can be expressed as
\begin{equation}\label{eq5}
	\Delta _\mathrm{t,SI}^{{l_\mathrm{t,SI}}}\left( {{\mathbf{t}_{{n_\mathrm{t}}}}} \right) = \frac{{2 \pi{\mathbf{s}^{{l_\mathrm{t,SI}}}}\left( {{\mathbf{t}_{n_\mathrm{t}}} - {\mathbf{o}_\mathrm{t}}} \right)}}{\lambda } ,
\end{equation}
where $\lambda$ is the wavelength. Thus, the field-response vector of the $n_\mathrm{t}$-th transmit MA and the field-response matrix of $N_\mathrm{t}$ transmit MAs are respectively given by
\begin{align}\label{eq6}
	&\mathbf{g}_\mathrm{SI}\left( {{\mathbf{t}_{{n_\mathrm{t}}}}} \right) = {\left[ {{e^{j\Delta _\mathrm{t,SI}^1\left( {{\mathbf{t}_{n_\mathrm{t}}}} \right)}}, \cdots ,{e^{j\Delta _\mathrm{t,SI}^{{L_\mathrm{t,SI}}}\left( {{\mathbf{t}_{{n_\mathrm{t}}}}} \right)}}} \right]^\mathrm{T}} \in {\mathbb{C}^{{L_\mathrm{t,SI}} \times 1}}, \\
	&\mathbf{G}_\mathrm{SI}\left( {\tilde{\mathbf{t}}} \right) = \left[ \mathbf{g}_\mathrm{SI}\left( {\mathbf{t}_1} \right),\mathbf{g}_\mathrm{SI}\left( {\mathbf{t}_2} \right), \cdots, \mathbf{g}_\mathrm{SI}\left( {\mathbf{t}_{N_\mathrm{t}}} \right) \right] \in {\mathbb{C}^{{L_\mathrm{t,SI}} \times N_\mathrm{t}}}.
\end{align}

Similarly, the field-response matrix of $N_\mathrm{r}$ receive MAs can be written as
\begin{equation}\label{eq8}
	\mathbf{F}_\mathrm{SI}\left( {\tilde{\mathbf{r}}} \right) = \left[ \mathbf{f}_\mathrm{SI}\left( {\mathbf{r}_1} \right),\mathbf{f}_\mathrm{SI}\left( {\mathbf{r}_2} \right), \cdots, \mathbf{f}_\mathrm{SI}\left( {\mathbf{r}_{N_\mathrm{r}}} \right) \right] \in {\mathbb{C}^{{L_\mathrm{r,SI}} \times N_\mathrm{r}}},
\end{equation}
where $\mathbf{f}_\mathrm{SI}\left( {\mathbf{r}_{n_\mathrm{r}}} \right) \in {\mathbb{C}^{{L_\mathrm{r,SI}} \times 1}}$ is the field-response vector of the $n_\mathrm{r}$-th receive MA.

As a result, the SI channel matrix is obtained as
\begin{equation}\label{eq9}
	{\mathbf{H}_\mathrm{SI}} \left( {\tilde{ \mathbf{t}}}, {\tilde{ \mathbf{r}}} \right) ={\mathbf{F}_\mathrm{SI}}{\left( {\tilde {\mathbf{r}}} \right)^\mathrm{H}}\mathbf{\Sigma} {\mathbf{G}_\mathrm{SI}}\left( {\tilde {\mathbf{t}}} \right) ,
\end{equation}
where $\mathbf{\Sigma} \in {\mathbb{C}^{{{L}_\mathrm{r,SI}} \times{{L}_\mathrm{t,SI}}}}$ is the channel response from the origin of the transmit region to that of the receive region.
\subsubsection{SIMO or MISO Channels for ${\mathbf{h}_\mathrm{UB}} \left( {\tilde{ \mathbf{r}}} \right)$, ${\mathbf{h}_\mathrm{BD}} \left( {\tilde{ \mathbf{t}}} \right)$, and ${\mathbf{h}_\mathrm{BE}} \left( {\tilde{ \mathbf{t}}} \right)$}
Since the UL channel can be regarded as a single input multiple output (SIMO) channel and the DL one is a multiple input single output (MISO) channel, the corresponding channel responses are given by
\begin{align}\label{eq10}
	&{\mathbf{h}_\mathrm{UB}}\left( {\tilde {\mathbf{r}}} \right) = {\mathbf{F}_\mathrm{UB}}{\left( {\tilde {\mathbf{r}}} \right)^\mathrm{H}}{\mathbf{g}_\mathrm{UB}}, \\
	&{\mathbf{h}_q}\left( {\tilde {\mathbf{t}}} \right) = {{\mathbf{G}_q}\left( {\tilde  {\mathbf{t}}} \right)}^\mathrm{H}{\mathbf{f}_q},\quad q \in \left\{ \mathrm{BD,BE} \right\},
\end{align}
where ${\mathbf{F}_\mathrm{UB}}{\left( {\tilde {\mathbf{r}}} \right)} \in {\mathbb{C}^{{L_\mathrm{r,UB}} \times {N_\mathrm{r}}}}$ and ${\mathbf{G}_q}{\left( {\tilde {\mathbf{t}}} \right)} \in {\mathbb{C}^{{L_{\mathrm{t},q}} \times {N_\mathrm{t}}}}$ are the field-response matrices for the UL-to-BS channel and the BS-to-DL/Eve channel, respectively.   ${L_\mathrm{r,UB}}$ and ${L_{\mathrm{t},q}}$ are the corresponding number of paths.  ${\mathbf{g}_\mathrm{UB}} \in {\mathbb{C}^{{L_{\mathrm{r,UB}}} \times 1}}$ and ${\mathbf{f}_q} \in {\mathbb{C}^{{L_{\mathrm{t},q}} \times 1}}$ represent the channel responses from the UL user to the receive origin and from the transmit origin to the DL user/Eve, respectively. The UL and DL field-response matrices are calculated by the method described before, and readers can also refer to \cite{ref_MA_Secure2}.

\subsection{Problem Formulation}
The receive signal-to-interference-plus-noise ratios (SINRs) of the UL and DL users are respectively given by
\begin{align}\label{eq12}
	{\gamma _{\mathrm{U}}} & = \frac{{{{\left| {\mathbf{w}_\mathrm{r}^\mathrm{H}{\mathbf{h}_\mathrm{UB}}\left( {\tilde {\mathbf{r}}} \right)} \right|}^2}{P_\mathrm{U}}}}{{{{\rho \left| {\mathbf{w}_\mathrm{r}^\mathrm{H}  {\mathbf{H}_\mathrm{SI}}\left( {\tilde { \mathbf{t}},\tilde { \mathbf{r}}} \right)\mathbf{x}} \right|}^2} + {{\left\| {\mathbf{w}_\mathrm{r}} \right\|}^2_2}\sigma _\mathrm{B}^2}}, \\
	{\gamma _\mathrm{D}} & = \frac{{{{\left| {\mathbf{h}_\mathrm{BD}^\mathrm{H}\left( {\tilde {\mathbf{t}}} \right)\mathbf{w}} \right|}^2}}}{{{{\left| {\mathbf{h}_\mathrm{BD}^\mathrm{H}\left( {\tilde {\mathbf{t}}} \right)\mathbf{v}} \right|}^2} + {{\left| {{h_\mathrm{UD}}} \right|}^2}{P_\mathrm{U}} + \sigma _\mathrm{D}^2}}.
\end{align}
Suppose that the Eve separately eavesdrops on the UL and DL transmissions \cite{ref10,ref11}.  Thus, the SINRs of the UL and DL interceptions at the Eve are respectively given by
\begin{align}\label{eq14}
	\gamma _\mathrm{U}^\mathrm{E} & = \frac{{{{\left| {{h_\mathrm{UE}}} \right|}^2}{P_\mathrm{U}}}}{{{{\left| {\mathbf{h}_\mathrm{BE}^\mathrm{H}\left( {\tilde {\mathbf{t}}} \right)\mathbf{x}} \right|}^2} + \sigma _\mathrm{E}^2}}, \\
	\gamma _\mathrm{D}^\mathrm{E} & = \frac{{{{\left| {\mathbf{h}_\mathrm{BE}^\mathrm{H}\left( {\tilde {\mathbf{t}}} \right)\mathbf{w}} \right|}^2}}}{{{{\left| {\mathbf{h}_\mathrm{BE}^\mathrm{H}\left( {\tilde {\mathbf{t}}} \right)\mathbf{v}} \right|}^2} + {{\left| {{h_\mathrm{UE}}} \right|}^2}{P_\mathrm{U}} + \sigma _\mathrm{E}^2}}.
\end{align}
Then, the achievable secrecy rates of the UL and DL users are $R_\mathrm{U}^{\sec } = {\left[ {{{\log }_2}\left( {1 + {\gamma _{\mathrm{U}}}} \right) - {{\log }_2}\left( {1 + \gamma _\mathrm{U}^\mathrm{E}} \right)} \right]^ + }$ and $R_\mathrm{D}^{\sec } = {\left[ {{{\log }_2}\left( {1 + {\gamma _{\mathrm{D}}}} \right) - {{\log }_2}\left( {1 + \gamma _\mathrm{D}^\mathrm{E}} \right)} \right]^ + }$, respectively\footnote{The operator ${\left[  \cdot  \right]^ + }$ has no impact on the optimization and is thus omitted in the subsequent derivations.}.

In this paper, we focus on maximizing the SSR of the users, i.e., $R^{\sec }={R_\mathrm{U}^{\sec } + R_\mathrm{D}^{\sec }}$, by jointly optimizing the beamformers, ${\mathbf{w}_\mathrm{r}}$, $\mathbf{w}$, and $\mathbf{v}$, and the MAs' positions of the BS, $\tilde{\mathbf{t}}$ and $\tilde{\mathbf{r}}$.  The optimization problem is formulated as
\begin{align}
	\mathop {\max }\limits_{{\mathbf{w}_\mathrm{r}},\mathbf{w},\mathbf{v},\tilde{\mathbf{t}},\tilde{\mathbf{r}}} \quad
	&{R^{\sec }}  \label{eq16}\\
	\mathrm{s.t.} \quad
	&\left\| {{\mathbf{w}_\mathrm{r}}} \right\|_2^2 = 1, \tag{15a} \label{16a}\\
	&\mathrm{Tr}\left( {\mathbf{ww}^\mathrm{H} + \mathbf{vv}^\mathrm{H}} \right) \le {P_\mathrm{B}}, \tag{15b} \label{16b}\\
	&\tilde {\mathbf{t}} \in {\mathcal{C}_\mathrm{t}}, \ \tilde {\mathbf{r}} \in {\mathcal{C}_\mathrm{r}}, \tag{15c} \label{16c}\\
	&{\left\| {{\mathbf{t}_a} - {\mathbf{t}_b}} \right\|_2} \ge D,1 \le a \ne b \le {N_\mathrm{t}}, \tag{15d} \label{16d}\\
	&{\left\| {{\mathbf{r}_a} - {\mathbf{r}_b}} \right\|_2} \ge D,1 \le a \ne b \le {N_\mathrm{r}}. \tag{15e} \label{16e}
\end{align}

Constraint \eqref{16b} is the maximum transmitted power ${P_\mathrm{B}}$ of the BS. Constraint \eqref{16c} limits the ranges of MAs' movements. Constraints \eqref{16d} and \eqref{16e} ensure that minimum inter-MA distance $D$ at the BS for practical implementation. Note that \eqref{eq16} is a non-convex optimization problem, which requires extremely high computational complexity to attain the globally optimal solution. Hence, we propose an AO method based on the successive convex approximation (SCA), semidefinite relaxation (SDR), and particle swarm optimization (PSO) in the next section.
\section{Proposed Solution}
In this section, we propose an AO method to decompose the original problem into three subproblems and solve them iteratively.
\subsection{Subproblem 1: Optimize $\mathbf{w}_\mathrm{r}$ With Given $\mathbf{w}$, $\mathbf{v}$, $\tilde {\mathbf{t}}$, and $\tilde {\mathbf{r}}$}
As the receive beamformer $\mathbf{w}_\mathrm{r}$ is only related to the UL SINR ${\gamma _{\mathrm{U}}}$, subproblem 1 can be expressed as
\begin{align}
	\mathop {\max }\limits_{{\mathbf{w}_\mathrm{r}}} \quad
	&\frac{{\mathbf{w}_\mathrm{r}^\mathrm{H}{\mathbf{h}_\mathrm{UB}}\left( {\tilde {\mathbf{r}}} \right)\mathbf{h}_\mathrm{UB}^\mathrm{H}\left( {\tilde {\mathbf{r}}} \right){\mathbf{w}_\mathrm{r}}}}{{\mathbf{w}_\mathrm{r}^\mathrm{H}\left[ {\rho {\mathbf{H}_\mathrm{SI}}\left( {\tilde {\mathbf{t}},\tilde {\mathbf{r}}} \right)\left( {\mathbf{W} + \mathbf{V}} \right)\mathbf{H}_\mathrm{SI}^\mathrm{H}\left( {\tilde {\mathbf{t}},\tilde {\mathbf{r}}} \right) + \sigma _\mathrm{B}^2{\mathbf{I}_{{N_\mathrm{r}}}}} \right]{\mathbf{w}_\mathrm{r}}}} \label{eq17}\\
	\mathrm{s.t.} \quad
	&\eqref{16a}, \tag{16a} \label{17a}
\end{align}
where $\mathbf{W}  = \mathbf{ww}^\mathrm{H} \in {\mathbb{C}^{{N_\mathrm{t}} \times {N_\mathrm{t}}}}$ and $\mathbf{V} = \mathbf{vv}^\mathrm{H} \in {\mathbb{C}^{{N_\mathrm{t}} \times {N_\mathrm{t}}}}$.

Let $\mathbf{A} \buildrel \Delta \over =  {\rho{\mathbf{H}_\mathrm{SI}}\left( {\tilde {\mathbf{t}},\tilde {\mathbf{r}}} \right)\left( {\mathbf{W} + \mathbf{V}} \right)\mathbf{H}_\mathrm{SI}^\mathrm{H}\left( {\tilde {\mathbf{t}},\tilde {\mathbf{r}}} \right) + \sigma _\mathrm{B}^2{\mathbf{I}_{{N_\mathrm{r}}}}} \in {\mathbb{C}^{{N_\mathrm{r}} \times {N_\mathrm{r}}}}$. The optimal solution of \eqref{eq17} can be obtained as
\begin{equation}\label{eq18}
	\mathbf{w}_\mathrm{r}^ *  = \frac{{{\mathbf{A}^{ - 1}}{\mathbf{h}_\mathrm{UB}}\left( {\tilde {\mathbf{r}}} \right)}}{{{{\left\| {{\mathbf{A}^{ - 1}}{\mathbf{h}_\mathrm{UB}}\left( {\tilde {\mathbf{r}}} \right)} \right\|}_2}}}.
\end{equation}
\subsection{Subproblem 2: Optimize $\mathbf{w}$ and $\mathbf{v}$ With Given $\mathbf{w}_\mathrm{r}$, $\tilde {\mathbf{t}}$, and $\tilde {\mathbf{r}}$}
Define ${h_1} \buildrel \Delta \over = \mathbf{w}_\mathrm{r}^\mathrm{H}{\mathbf{h}_\mathrm{UB}}\left( {\tilde {\mathbf{r}}} \right)\in \mathbb{C}^{1 \times 1}$, ${\mathbf{h}_2} \buildrel \Delta \over = \sqrt \rho  \mathbf{H}_\mathrm{SI}^\mathrm{H}\left( {\tilde {\mathbf{t}},\tilde {\mathbf{r}}} \right){\mathbf{w}_\mathrm{r}}\in {\mathbb{C}^{{N_\mathrm{t}} \times 1}}$, ${\mathbf{H}_2} \buildrel \Delta \over = {\mathbf{h}_2}\mathbf{h}_2^\mathrm{H}\in {\mathbb{C}^{{N_\mathrm{t}} \times {N_\mathrm{t}}}}$, ${\mathbf{H}_\mathrm{BD}} \buildrel \Delta \over = {\mathbf{h}_\mathrm{BD}\left( {\tilde {\mathbf{t}}} \right)}\mathbf{h}_\mathrm{BD}^\mathrm{H}\left( {\tilde {\mathbf{t}}} \right)\in {\mathbb{C}^{{N_\mathrm{t}} \times {N_\mathrm{t}}}}$, and ${\mathbf{H}_\mathrm{BE}} \buildrel \Delta \over = {\mathbf{h}_\mathrm{BE}\left( {\tilde {\mathbf{t}}} \right)}\mathbf{h}_\mathrm{BE}^\mathrm{H}\left( {\tilde {\mathbf{t}}} \right)\in {\mathbb{C}^{{N_\mathrm{t}} \times {N_\mathrm{t}}}}$.  With the rule of logarithmic function, we rewrite the SSR as
\begin{equation}\label{eq19}
	{R^{\sec }} = \mathcal{P}\left( {\mathbf{W},\mathbf{V}} \right) - \mathcal{Q}\left( {\mathbf{W},\mathbf{V}} \right),
\end{equation}
where
\begin{align}\label{eq20}
	&\mathcal{P}\left( {\mathbf{W},\mathbf{V}} \right)
	 = {\log _2}\left\{ {\mathrm{Tr}\left[ {\left( \mathbf{W + V} \right){\mathbf{H}_2}} \right] + {P_\mathrm{U}}{{\left| {{h_1}} \right|}^2} + \sigma _\mathrm{B}^2} \right\} \nonumber\\
	&\qquad + {\log _2}\left\{ {\mathrm{Tr}\left[ {\left( \mathbf{W + V} \right){\mathbf{H}_\mathrm{BD}}} \right] + {P_\mathrm{U}}{{\left| {{h_\mathrm{UD}}} \right|}^2} + \sigma _\mathrm{D}^2} \right\} \nonumber\\
	&\qquad + {\log _2}\left\{ {\mathrm{Tr}\left[ {\left( \mathbf{W + V} \right){\mathbf{H}_\mathrm{BE}}} \right] + \sigma _\mathrm{E}^2} \right\} \nonumber\\
	&\qquad + {\log _2}\left[ {\mathrm{Tr} {\left( \mathbf{V} {\mathbf{H}_\mathrm{BE}}\right)} + {P_\mathrm{U}}{{\left| {{h_\mathrm{UE}}} \right|}^2} + \sigma _\mathrm{E}^2} \right], \\
	&\mathcal{Q}\left( {\mathbf{W},\mathbf{V}} \right)
	= {\log _2}\left\{ {\mathrm{Tr}\left[ {\left( \mathbf{W + V} \right){\mathbf{H}_2}} \right] + \sigma _\mathrm{B}^2} \right\} \nonumber\\
	&\qquad + {\log _2}\left[ {\mathrm{Tr}\left( {\mathbf{V}{\mathbf{H}_\mathrm{BD}}} \right) + {P_\mathrm{U}}{{\left| {{h_\mathrm{UD}}} \right|}^2} + \sigma _\mathrm{D}^2} \right] \nonumber\\
	&\qquad + 2{\log _2}\left\{ {\mathrm{Tr}\left[ {\left( \mathbf{W + V} \right){\mathbf{H}_\mathrm{BE}}} \right] + {P_\mathrm{U}}{{\left| {{h_\mathrm{UE}}} \right|}^2} + \sigma _\mathrm{E}^2} \right\}.
\end{align}
After that, subproblem 2 can be formulated as
\begin{align}
	\mathop {\max }\limits_{\mathbf{W},\mathbf{V}} \quad
	&\mathcal{P}\left( {\mathbf{W},\mathbf{V}} \right) - \mathcal{Q}\left( {\mathbf{W},\mathbf{V}} \right) \label{eq22}\\
	\mathrm{s.t.} \quad
	&\mathrm{Tr}\left( \mathbf{W+V} \right) \le P_\mathrm{B}, \tag{21a} \label{22a}\\
	&\mathbf{W} \succeq \mathbf{0}, \ \mathbf{V} \succeq \mathbf{0}, \tag{21b} \label{22b}\\
	&\mathrm{rank}\left( \mathbf{W} \right) \le 1, \ \mathrm{rank}\left( \mathbf{V} \right) \le 1. \tag{21c} \label{22c}
\end{align}

Note that both $\mathcal{P}\left( {\mathbf{W},\mathbf{V}} \right)$ and $\mathcal{Q}\left( {\mathbf{W},\mathbf{V}} \right)$ are concave functions, and thus the objective function in \eqref{eq22} is a difference-of-convex function. Therefore, the SCA is applied to obtain a locally optimal solution \cite{ref14}. Specifically, take the first-order Taylor expansion as a global overestimate of the differentiable concave function $\mathcal{Q}\left( {\mathbf{W},\mathbf{V}} \right)$ around the local point $\left( {{\mathbf{W}^{\left( {m - 1} \right)}},{\mathbf{V}^{\left( {m - 1} \right)}}} \right)$ with $1 \le m \le M$, where $M$ is the maximum number of iterations for SCA, i.e.,
\begin{align}\label{23}
	& \mathcal{Q}\left( {\mathbf{W},\mathbf{V}\left| {{\mathbf{W}^{\left( {m - 1} \right)}},{\mathbf{V}^{\left( {m - 1} \right)}}} \right.} \right) = \mathcal{Q}\left( {{\mathbf{W}^{\left( {m - 1} \right)}},{\mathbf{V}^{\left( {m - 1} \right)}}} \right) \nonumber\\
	& + \mathrm{Tr}\left[ {{{\left( {{\nabla _\mathbf{W}}\mathcal{Q}\left( {{\mathbf{W}^{\left( {m - 1} \right)}},{\mathbf{V}^{\left( {m - 1} \right)}}} \right)} \right)}^\mathrm{H}}\left( {\mathbf{W} - {\mathbf{W}^{\left( {m - 1} \right)}}} \right)} \right] \nonumber\\
	& + \mathrm{Tr}\left[ {{{\left( {{\nabla _\mathbf{V}}\mathcal{Q}\left( {{\mathbf{W}^{\left( {m - 1} \right)}},{\mathbf{V}^{\left( {m - 1} \right)}}} \right)} \right)}^\mathrm{H}}\left( {\mathbf{V} - {\mathbf{V}^{\left( {m - 1} \right)}}} \right)} \right] \nonumber\\
	& \ge  \mathcal{Q}\left( \mathbf{W,V} \right).
\end{align}

Then, after relaxing the rank constraint \eqref{22c} by removing it, the following convex problem can serve as a lower bound of \eqref{eq22}.
\begin{align}
	\mathop {\max }\limits_{\mathbf{W},\mathbf{V}} \quad
	&\mathcal{P}\left( {\mathbf{W},\mathbf{V}} \right) - \mathcal{Q}\left( {\mathbf{W},\mathbf{V}\left| {{\mathbf{W}^{\left( {m - 1} \right)}},{\mathbf{V}^{\left( {m - 1} \right)}}} \right.} \right) \label{eq24}\\
	\mathrm{s.t.} \quad
	&{\eqref{22a}-\eqref{22b}}. \tag{23a} \label{24a}
\end{align}
By iteratively solving \eqref{eq24} with the aid of the CVX toolbox, the objective value will increase and converge \cite{ref14}. When the increment of the objective value between two iterations is less than $\varepsilon_1$, the optimized $\mathbf{W}^*$ and $\mathbf{V}^*$ are outputted. Besides, the tightness of the rank relaxation is verified in the \cite{ref10, ref11}. Thus, in this case, the optimized beamformers $\mathbf{w}^*$ and $\mathbf{v}^*$ are obtained by the eigenvalue decomposition.
\subsection{Subproblem 3: Optimize $\tilde {\mathbf{t}}$ and $\tilde {\mathbf{r}}$ With Given $\mathbf{w}_\mathrm{r}$, $\mathbf{w}$, and $\mathbf{v}$}
With the given $\mathbf{w}_\mathrm{r}$, $\mathbf{w}$, and $\mathbf{v}$, the SSR can be expressed as a function of $\tilde {\mathbf{t}}$ and $\tilde {\mathbf{r}}$.  Therefore, subproblem 3 can be formulated as
\begin{align}
	\mathop {\max }\limits_{\tilde{\mathbf{t}},\tilde{\mathbf{r}}} \quad
	&{R^{\sec } \left( \tilde{\mathbf{t}}, \tilde{\mathbf{r}} \right)}  \label{eq25}\\
	\mathrm{s.t.} \quad
	&{\eqref{16c}-\eqref{16e}}. \tag{24a} \label{25a}
\end{align}

The conventional alternating position optimization, which alternates by fixing the other MAs and moving only one of them, may converge to an undesired local optimal solution. Therefore, the PSO \cite{ref12} is employed to simultaneously optimize the positions of all transmit MAs or all receive MAs.

The positions of all transmit MAs are first optimized with given $\tilde {\mathbf{r}}$ by the following steps. First, initialize the positions and velocities of particles as $\tilde {\mathbf{t}}_i^{\left( 0 \right)}$ and $\mathbf{p}_i^{\left( 0 \right)}$, respectively, with $1 \le i \le I$, where $I$ is the number of particles. Second, the individual best position of the $i$-th particle $\tilde{\mathbf{t}}_i^*$ and the global best position $\tilde{\mathbf{t}}^*$ are selected based on the fitness function. Then, the velocity and position of each particle at the $k$-th iteration with $1 \le k \le K$, where $K$ is the maximum number of iterations for PSO, are updated as
\begin{align}
	\mathbf{p}_i^{\left( k \right)} &= \omega \mathbf{p}_i^{\left( {k - 1} \right)} + {c_1}{\mathbf{e}_1} \odot \left( {\tilde {\mathbf{t}}_i^* - \tilde {\mathbf{t}}_i^{\left( {k - 1} \right)}} \right) \nonumber \\
	&\qquad\qquad\qquad + {c_2}{\mathbf{e}_2} \odot \left( {{{\tilde {\mathbf{t}}}^*} - \tilde {\mathbf{t}}_i^{\left( {k - 1} \right)}} \right), \\
	\tilde {\mathbf{t}}_i^{\left( k \right)} &= \mathcal{B}\left\{ {\tilde {\mathbf{t}}_i^{\left( {k - 1} \right)} + \mathbf{p}_i^{\left( k \right)}} \right\},
\end{align}
where $\omega$ is a linear function decreasing with the number of iterations in the interval $\left[\omega_\mathrm{min},\omega_\mathrm{max}\right] $, i.e., $\omega = \omega_\mathrm{max} - \left( \omega_\mathrm{max} - \omega_\mathrm{min} \right) k/K$. $c_1$ and $c_2$ are the individual and global learning factors that push each particle toward the individual and global best positions, respectively. Two random vectors $\mathbf{e}_1$ and $\mathbf{e}_2$, where each entry is a uniform random number in the range $\left[0,1\right]$, are utilized to increase the randomness for reducing the possibility of converging to an undesired local optimal solution. $\mathcal{B} \left\lbrace \mathbf{a}\right\rbrace $ is a function that projects each entry of the vector $\mathbf{a}$ to its corresponding maximum/minimum value to satisfy constraint \eqref{16c}.

During each iteration, the individual and global best positions are updated according to the fitness function. Assume that the best position has the largest fitness value. Considering the constraints \eqref{16d} and \eqref{16e} on the inter-MA distance, we add a penalty term to the fitness function. For maximizing the SSR, the fitness function is defined as
\begin{equation}\label{eq28}
	\mathcal{F}\left( {\tilde {\mathbf{t}}_i^{\left( k \right)}} \right) = {R^{\sec }}\left( {\tilde {\mathbf{t}}_i^{\left( k \right)}} \right) - \eta  \Gamma \left\{ {\tilde {\mathbf{t}}_i^{\left( k \right)}} \right\},
\end{equation}
where $\Gamma \left\{ {\tilde {\mathbf{t}}} \right\}$ is a function that returns the number of MAs that violate the minimum inter-MA distance constraint at position ${\tilde {\mathbf{t}}}$. $\eta$ is a large positive penalty factor to consistently hold ${R^{\sec }}\left( {\tilde {\mathbf{t}}} \right)-\eta \le 0$. Therefore, this penalty term can push the particles to satisfy the minimum inter-MA distance. Finally, after $K$ iterations, the positions of all transmit MAs $\tilde {\mathbf{t}}^*$ are suboptimal in general.

When the optimized positions of the transmit MAs are given, the corresponding positions of the receive MAs $\tilde {\mathbf{r}}^*$ are also obtained via the PSO. Note that the PSO-based receive MAs' positions optimization has similar steps as obtaining $\tilde {\mathbf{t}}^*$.  Furthermore, we can divide the $N_\mathrm{t}+N_\mathrm{r}$ MAs into $\delta$ sets ($1 \le \delta \le N_\mathrm{t}+N_\mathrm{r}$) to optimize their positions separately. In general, as the number of MAs in a set increases, it is more difficult to simultaneously make all MAs satisfy the minimum inter-MA distance constraint. Conversely, it is easier to converge to an undesired suboptimal solution. How to balance this trade-off is left as a potential issue for future work.

\subsection{Overall Algorithm and Analysis}
As has been discussed, the SSR maximization can be decomposed into three subproblems, and the corresponding algorithms are proposed to solve them.  The overall algorithm is shown in Algorithm \ref{alg1}, where $\varepsilon_2 > 0$ denotes a small threshold and $C$ is the maximum number of iterations for AO.
\begin{algorithm}[!t]
	\caption{Alternating Optimization Method}
	\label{alg1}
	\renewcommand{\algorithmicrequire}{\textbf{Initialization:}}
	\renewcommand{\algorithmicensure}{\textbf{Output:}}
	\begin{algorithmic}[1]
		\REQUIRE Set initial PSO parameters, $\mathbf{w}^{\left( 0 \right) }$, $\mathbf{v}^{\left( 0 \right) }$, $\mathbf{w}_\mathrm{r}^{\left( 0 \right) }$, and the index of AO $c=0$.
		\ENSURE $\tilde {\mathbf{t}}$, $\tilde {\mathbf{r}}$, ${\mathbf{w}}$, ${\mathbf{v}}$, and ${\mathbf{w}}_\mathrm{r}$.
		\STATE Calculate the value of the initial SSR $R^{\sec \left(  0 \right)}$;
		\REPEAT
		\STATE Update $c=c+1$;
		\STATE With given $\mathbf{w}^{\left( c-1 \right) }$, $\mathbf{v}^{\left( c-1 \right) }$, and $\mathbf{w}_\mathrm{r}^{\left( c-1 \right) }$, solve subproblem 3 and store the intermediate solutions $\tilde {\mathbf{t}}^{\left( c \right) }$ and $\tilde {\mathbf{r}}^{\left( c \right) }$;
		\STATE With given $\tilde {\mathbf{t}}^{\left( c \right) }$, and $\tilde {\mathbf{r}}^{\left( c \right) }$, and $\mathbf{w}_\mathrm{r}^{\left( c-1 \right) }$, solve subproblem 2 and store the intermediate solutions $\mathbf{w}^{\left( c \right) }$ and $\mathbf{v}^{\left( c \right) }$;
		\STATE With given $\tilde {\mathbf{t}}^{\left( c \right) }$, $\tilde {\mathbf{r}}^{\left( c \right) }$, $\mathbf{w}^{\left( c \right) }$, and $\mathbf{v}^{\left( c \right) }$, update $\mathbf{w}_\mathrm{r}^{\left( c \right) }$ according to \eqref{eq18};
		\STATE Update the SSR $R^{\sec \left(  c \right)}$;
		\UNTIL {$\frac{{{R^{\sec }}^{\left( c \right)} - {R^{\sec }}^{\left( {c - 1} \right)}}}{{{R^{\sec }}^{\left( c \right)}}} \le \varepsilon_2$ or $c \ge C$;}
		\RETURN $\tilde {\mathbf{t}} = \tilde {\mathbf{t}}^{\left( c \right) }$, $\tilde {\mathbf{r}} = \tilde {\mathbf{r}}^{\left( c \right) }$, ${\mathbf{w}}=\mathbf{w}^{\left( c \right) }$, ${\mathbf{v}}=\mathbf{v}^{\left( c \right) }$, and ${\mathbf{w}}_\mathrm{r}=\mathbf{w}_\mathrm{r}^{\left( c \right) }$.
	\end{algorithmic}
\end{algorithm}

The convergence of Algorithm \ref{alg1} is guaranteed because $R^{\sec \left(  c \right)}$ is non-decreasing over iterations \cite{ref10} and is bounded with the limited communication resources.  The complexity of steps 4, 5, and 6 in Algorithm \ref{alg1} are $\mathcal{O}\left(2IK L_\mathrm{sum} \right) $, $\mathcal{O}\left(2M_\mathrm{SCA} N_\mathrm{t}^{3.5} \right) $, and $\mathcal{O}\left( N_\mathrm{r}^3 \right) $, respectively \cite{ref10}, where $L_\mathrm{sum} = \sum\nolimits_{i \in \left\{ \mathrm{SI,BD,BE} \right\}} {{{L}_{\mathrm{t},i}}}  + \sum\nolimits_{i \in \left\{ \mathrm{SI,UB} \right\}} {{{L}_{\mathrm{r},i}}} $, $M_\mathrm{SCA}$ is the number of iterations for SCA. Consequently, the complexity of the overall algorithm is $\mathcal{O}\left(C_\mathrm{AO} \left(2IKL_\mathrm{sum} +2M_\mathrm{SCA} N_\mathrm{t}^{3.5} +N_\mathrm{r}^3\right) \right) $, where $C_\mathrm{AO}$ denotes the number of iterations for AO.
\section{Simulation Results}
\begin{table}[!t]
	\caption{Simulation Parameters}
	\label{tab1}
	\centering
	\begin{tabular}{|l|l|l|}
		\hline
		\multicolumn{1}{|c|}{\textbf{Parameter}} & \multicolumn{1}{c|}{\textbf{Description}} & \multicolumn{1}{c|}{\textbf{Value}} \\ \hline
		$N$ & Number of transmit/receive MAs & 4 \\ \hline
		$A$ & Length of the sides of moving regions  & $2\lambda$ \\ \hline
		$L$ & Number of transmit/receive paths  & $3$ \\ \hline
		$D$ & Minimum inter-MA distance & $\lambda/2$ \\ \hline
		$\beta$ & Path loss of 1 meter  & -40dB   \\ \hline
		$\alpha$ & Path loss exponent  & 2.8   \\ \hline
		$\sigma _\mathrm{B}^2$, $\sigma _\mathrm{D}^2$, $\sigma _\mathrm{E}^2$ & Average noise powers  & -90dBm   \\ \hline
		$\rho$ & SIC coefficient  & -100dB   \\ \hline
		$P_\mathrm{B}$ & Maximum transmitted power of the BS  & 20dBm   \\ \hline
		$P_\mathrm{U}$ & Transmitted power of the UL user  & 20dBm  \\ \hline
		$\varepsilon_1$, $\varepsilon_2$ & Convergence thresholds  & $10^{-3}$   \\ \hline
		$I$ & Number of particles  & 100   \\ \hline
		$M$, $K$, $C$ & Maximum number of iterations  & 100   \\ \hline
		$\eta$ & Penalty factor  & 100   \\ \hline
		$\omega_\mathrm{min}/\omega_\mathrm{max}$ & Minimum/Maximum weight  & 0.4/0.9   \\ \hline
		$c_1$, $c_2$ & Individual and global learning factors  & 1.4   \\ \hline
	\end{tabular}
\end{table}
In this section, we evaluate the performance of our proposed scheme. The default simulation parameters are shown in Table\;\ref{tab1}. Specifically, the moving regions are squares with size $A \times A$. The UL user, DL user, and Eve are randomly and uniformly distributed in a cell centered on the BS with a radius of 50 meters. The geometry channel model is adopted \cite{ref_MA_Modeling}, in which the number of all paths is equal to $L$. In addition, each channel element from the transmit reference point to the receive reference point is the independent and identically distributed (i.i.d.) CSCG variable following $\mathcal{CN}\left( 0, \beta d^{-\alpha} / L\right) $, where $d$ is the propagation distance. The elevation and azimuth angles of the transmit and receive paths are also the i.i.d. variables following the uniform distribution over $\left[ 0,\pi\right] $. Without loss of generality, we assume that the BS has the same number of transmit and receive antennas, i.e., $N_\mathrm{t}=N_\mathrm{r} \buildrel \Delta \over = N$. Moreover, the proposed PSO-based MA-assisted FD system is referred to as \textbf{MA-FD-PSO}, and the following four benchmark schemes are simulated for comparison. \textbf{MA-FD-PSO-NoAN} discards AN in order to investigate the impact of AN on the SSR. \textbf{FPA-FD} replaces MAs with FPAs to investigate the performance gain of MA. \textbf{MA-FD-RP} adopts the MA with random position (RP) to demonstrate the performance of the PSO-based position optimization. \textbf{MA-HD-PSO} adjusts the BS to the time-division HD mode to present the FD gain.
\begin{figure}[!t]
	\centering
	\subfloat[$L=3$]{\label{fig2a}\includegraphics[width=0.5\columnwidth]{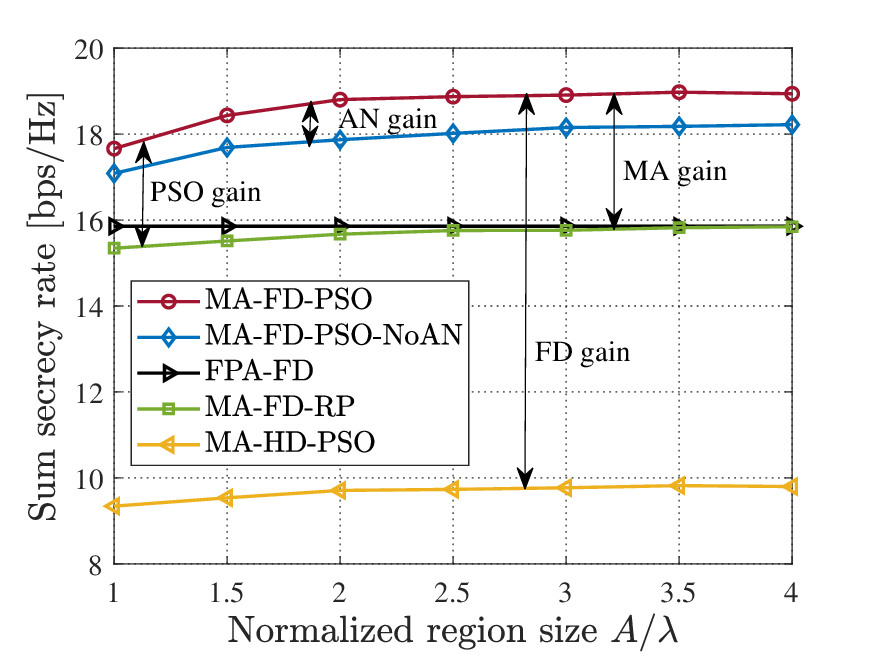}}
	\subfloat[$L=6$]{\label{fig2b}\includegraphics[width=0.5\columnwidth]{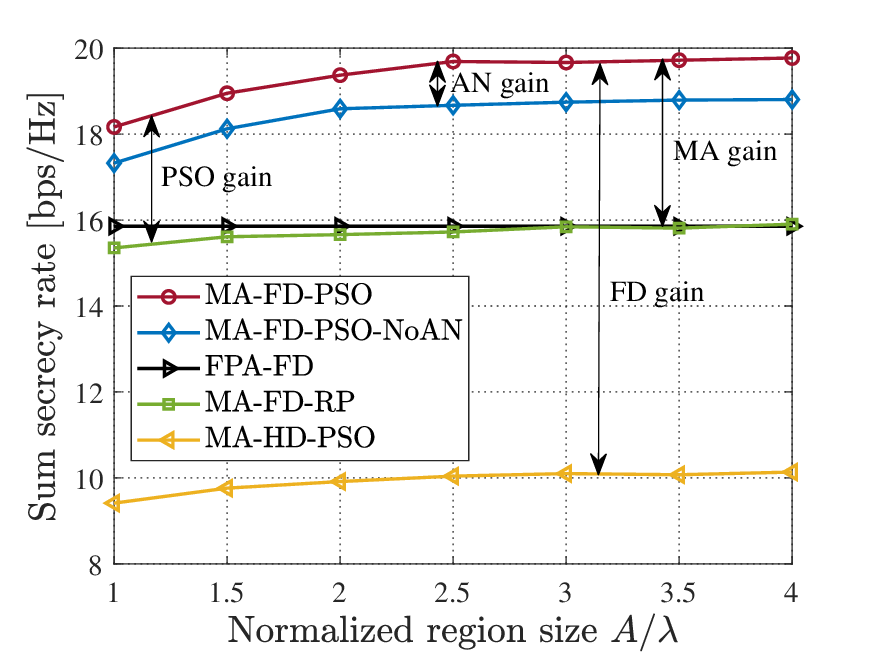}}\\
	\caption{Sum secrecy rate versus normalized region size.}
	\label{fig2}
\end{figure}
\begin{figure}[!t]
	\centering
	\includegraphics[width=0.68\columnwidth]{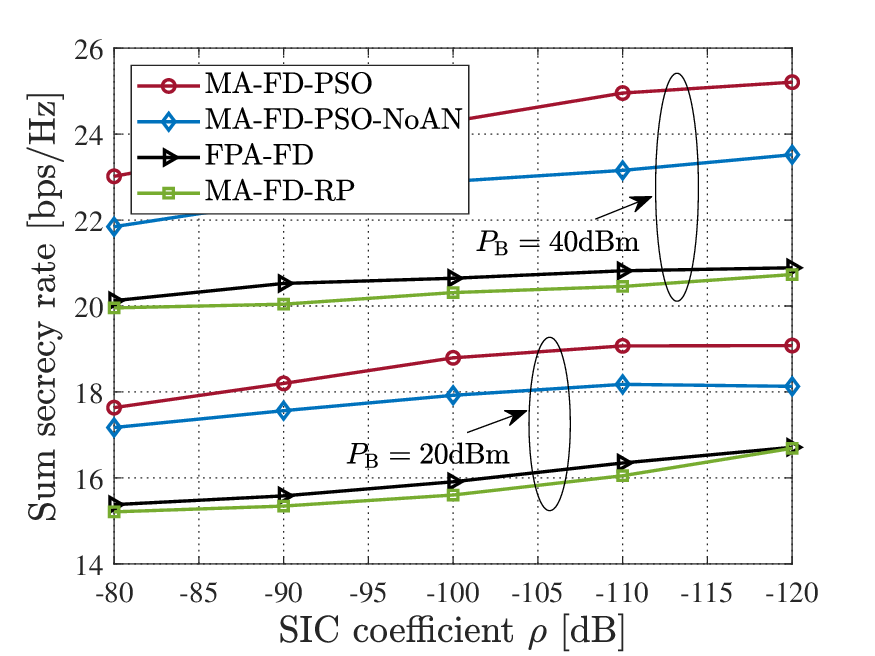}
	\caption{Sum secrecy rate versus the SIC coefficient.}
	\label{fig3}
\end{figure}
\begin{figure}[!t]
	\centering
	\includegraphics[width=0.68\columnwidth]{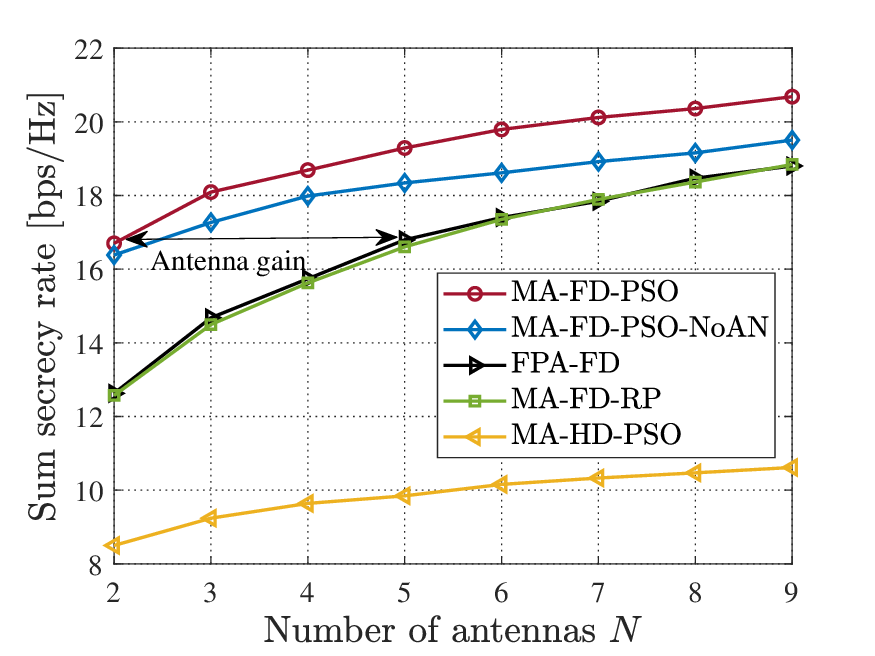}
	\caption{Sum secrecy rate versus the number of antennas.}
	\label{fig4}
\end{figure}

Fig. \ref{fig2} shows the SSR versus normalized region size for $L=3$ and $L=6$. The SSRs of all MA-based schemes increase with normalized region size $A/\lambda$ and the number of paths $L$ because more spatial DoFs can be explored with larger moving regions and more diversity gains can be obtained with more paths.  In Fig.\;\ref{fig2}\subref{fig2a}, the proposed MA-FD-PSO scheme has 2.32 bps/Hz PSO gain, 0.79 bps/Hz AN gain, 9.14 bps/Hz FD gain, and 3.69 bps/Hz MA gain compared to the benchmark schemes, which demonstrate the superiority of the proposed scheme. These four gains are also observed in Fig.\;\ref{fig2}\subref{fig2b}.

Fig. \ref{fig3} depicts the SSR versus the SIC coefficient under different maximum transmitted power of the BS. We can see that the SSRs of FD-based schemes rise with the SIC capability and BS's transmitted power. Among them, the proposed MA-FD-PSO scheme always has the highest SSR, which reveals the improvements brought by adjusting the positions of MAs, enabling FD mode, and generating the AN.

Fig. \ref{fig4} illustrates the SSR versus the number of transmit/receive antennas at the BS. By increasing the number of antennas, the spatial diversity is augmented and the beamforming performance can be improved, which lead to the enhancement of the SSR. In addition, the real-time antenna movement of the MA-based schemes enables further utilization of the spatial DoFs, and thus reduces the number of antennas at the same SSR level.  In this figure, the MA-FD-PSO scheme saves $2\times \left( 5-2\right) = 6$ antennas at an SSR threshold of 16.70 bps/Hz compared to the FPA-FD scheme.

\section{Conclusion}
This paper proposed a secure FD communication system aided by MA, in which the AN was utilized to enhance the performance of the PLS.  The beamformers of the BS and the positions of MAs were jointly optimized for the purpose of maximizing the SSR.  However, owing to the non-convex nature of the optimization problem, it was decomposed into three subproblems, which could be solved by the AO method based on the SCA, SDP, and PSO algorithms.  Simulation results substantiated that our proposed scheme could effectively improve the SSR in comparison to the four conventional schemes.

\end{document}